\documentclass[11pt, a4paper]{article}
\usepackage[utf8]{inputenc}
\usepackage[T1]{fontenc}
\usepackage{geometry}
\usepackage{amsmath}
\usepackage{amssymb}
\usepackage{float}
\usepackage{booktabs} 
\usepackage{hyperref} 
\usepackage{enumitem} 
\usepackage{titlesec} 
\usepackage{graphicx} 
\usepackage{multirow} 
\usepackage{subcaption} 
\usepackage{color}
\usepackage{makecell}
\usepackage[square,numbers,sort&compress]{natbib}
\title{\textbf{Fold-CP: A Context Parallelism Framework for Biomolecular Modeling}}
\author{
Dejun Lin\textsuperscript{1,*} \, 
Simon Chu\textsuperscript{1,*} \,
Vishanth Iyer \textsuperscript{1,*} \, 
Youhan Lee \textsuperscript{1,*} \,
John St John\textsuperscript{1,*} \,
\\
Kevin Boyd\textsuperscript{1,*} \, 
Brian Roland\textsuperscript{1,*} \, 
Xiaowei Ren\textsuperscript{1,*} \, 
Guoqing Zhou\textsuperscript{1,*} \,
Zhonglin Cao\textsuperscript{1,*} \,
\\
Polina Binder\textsuperscript{1,*} \, 
Yuliya Zhautouskaya\textsuperscript{1,*} \, 
Jakub Zakrzewski\textsuperscript{1,*} \,
Maximilian Stadler\textsuperscript{1,*} \,
\\
Kyle Gion\textsuperscript{1} \,
Yuxing Peng \textsuperscript{1} \,
Xi Chen\textsuperscript{1} \,
Tianjing Zhang\textsuperscript{1} \,
\\
Philipp Junk\textsuperscript{2} \,
Michelle Dimon\textsuperscript{2} \,
Pawe{\l} Gniewek\textsuperscript{2} \,
Fabian Ortega\textsuperscript{2} \,\\
McKinley Polen\textsuperscript{2} \,
Ivan Grubisic\textsuperscript{2} \,
Ali Bashir\textsuperscript{2} \,
\\
Graham Holt\textsuperscript{3} \,
Danny Kovtun\textsuperscript{3} \,
Matthias Grass\textsuperscript{3} \,
Luca Naef\textsuperscript{3} \,
\\
Rui Wang\textsuperscript{4} \,
Jian Peng\textsuperscript{4} \,
\\
Anthony Costa\textsuperscript{1} \,
Saee Paliwal\textsuperscript{1} \, 
Eddie Calleja\textsuperscript{1} \, 
Timur Rvachov\textsuperscript{1} \, 
\\
Neha Tadimeti\textsuperscript{1,*} \,
Roy Tal\textsuperscript{1,*} \,
Emine Kucukbenli\textsuperscript{1,*, \dag} 
\\
\\
{\small \textsuperscript{1}NVIDIA \, \textsuperscript{2}Rezo Therapeutics \,
\textsuperscript{3}Proxima \,
\textsuperscript{4}Earendil Labs \,}
\\
{\small \textsuperscript{*}Core contributor \, 
\textsuperscript{\dag} Correspondence to: ekucukbenli@nvidia.com \,}
}

\date{March 2026}

\begin{document}
\maketitle

\begin{abstract}
 
Understanding cellular machinery requires atomic-scale reconstruction of large biomolecular assemblies.
However, predicting the structures of these systems has been constrained by hardware memory requirements of models like AlphaFold 3, imposing a practical ceiling of a few thousand residues that can be processed on a single GPU. 
Here we present NVIDIA BioNeMo Fold-CP, a context parallelism framework that overcomes this barrier by distributing the inference and training pipelines of co-folding models across multiple GPUs. We use the Boltz models as open source reference architectures and implement custom multi-dimensional primitives that efficiently parallelize both the dense triangular updates and the irregular, data-dependent pattern of window-batched local attention.
Our approach achieves efficient memory scaling; for an $N$-token input distributed across $P$ GPUs, per-device memory scales as $O(N^2/P)$, enabling the structure prediction of assemblies exceeding 30,000 residues on 64 NVIDIA B300 GPUs.
We demonstrate the scientific utility of this approach through successful developer use cases: Fold-CP enabled the scoring of over 90\% of Comprehensive Resource of Mammalian protein complexes (CORUM) database, as well as folding of disease-relevant PI4KA lipid kinase complex bound to an intrinsically disordered region without cropping. By providing a scalable pathway for modeling massive systems with full global context, Fold-CP represents a significant step toward the realization of a virtual cell.
\end{abstract}

\section{Introduction}

Accurate folding of biomolecules is critical to rational drug design, since biological function and therapeutic viability often depend on the 3D structure. Historically, structure determination relied on experimental techniques such as X-ray crystallography. While these methods provide high-resolution insights, they are limited by cost and throughput. The advent of 3D structure prediction models like AlphaFold 2 (AF2)\cite{alphafold2} and AlphaFold-Multimer (AF-MM)\cite{afmm}  has transformed this landscape by enabling accurate prediction of small monomeric and multimeric proteins.

However, biological function and disease often arise from the concerted interactions of massive macromolecular assemblies. 
AlphaFold 3 (AF3)\cite{alphafold3} has already pushed the boundaries of these models by co-folding across biomolecules, such as proteins, DNA, RNA, and small molecules. The next frontier of research remains as transitioning from modeling isolated small proteins to larger macromolecular structures and interfaces. For instance, modeling massive biomolecules, such as viral capsids that frequently exceed 15,000 tokens, is critical because they are essential vectors for gene therapy and can be optimized with machine learning \cite{Tan2025}. 


Despite their success, AF3-like models hit a ceiling when it comes to large biomolecular systems. These models use a memory intensive, dense, pairwise representation to introduce physics-inspired attention mechanism in the neural network architecture. The memory requirement for this representation scales quadratically ($O(N^2)$) with sequence length, while the computational cost of the geometric operations scales cubically ($O(N^3)$), where $N$ is the number of input tokens. On standard infrastructure this imposes a strict practical limit of a few single-digit thousand tokens, for instance Boltz-1 reports a 2048‑token sequence cap on an NVIDIA A100 80GB GPU \cite{boltz1}. Consequently, over 70\% of characterized mammalian protein complexes, such as those cataloged in the CORUM database \cite{corum2024}, exceed single-GPU memory capacity. To circumvent this, current methods rely on artificial sequence cropping (which severs global context and introduces fragmentation bias), serial chunking (which causes latency bottlenecks), or architectural approximations like linear attention (which approximates the pairwise interaction) \cite{seedfold}.

Here we introduce NVIDIA BioNeMo \textbf{Fold-CP}, a Context Parallelism (CP) Framework for geometric co-folding models, designed to address the memory limitations without resorting to attention approximations or serial chunking. 
Using the Boltz architectures \cite{boltz1, boltz2} as a reference, Fold-CP re-architects the distributed memory hierarchy. 
Inspired by parallelism in Large Language Model (LLM) literature \cite{megatron-lm, liu2023ringattentionblockwisetransformers}, yet grounded in the constraints of these models,  
Fold-CP implements a 3D device mesh, comprising one Data Parallelism (DP) dimension and a 2D Context Parallelism (CP) grid, which collectively shards activations along the sequence, MSA, and atom dimensions.
In order to mitigate communication overhead, Fold-CP introduces novel distributed algorithms for triangle attention, triangle multiplication, pair weighted averaging, outer product mean, attention pair bias, and window batching.
Thanks to efficient orchestration of inter-process communication alongside process-local computation, Fold-CP enables both training and inference in unprecedented context lengths, while maintaining accuracy parity with the single device baseline. Beyond computational benchmarks, we demonstrate the scientific utility of Fold-CP through successful use cases and open questions regarding model behavior on large complexes. 
Ultimately, by providing the engine to scale context beyond a single device, Fold-CP establishes the foundational infrastructure necessary to
help unlock the next scale of biological modeling.

\subsection{Related Work}

The necessity to circumvent hardware memory limitations during the training and inference of structure prediction models is not a new challenge. Initial efforts primarily targeted high-throughput execution. Models like ParaFold \cite{parafold} and APACE \cite{apace} optimized the pipeline by decoupling CPU-heavy tasks such as Multiple Sequence Alignment (MSA) and template generation from GPU inference. While this prevented repeated recompilation and enabled the efficient batched execution needed to build conformational ensembles, it did not resolve the memory bottleneck for processing a single, massive assembly.

The development of optimized kernels like cuEquivariance \cite{cuequivariance} has partially mitigated these hardware constraints by reducing the memory footprint of triangular operations from cubic to quadratic, allowing current structure prediction models to infer on moderately larger biomolecular structures before hitting the limits of single-device memory.

To directly address the activation memory wall within AF2 \cite{alphafold2}, FastFold \cite{fastfold} introduced Dynamic Axial Parallelism (DAP), a sequence-parallel strategy that partitions activations across one of the sequential axes of the 2D pair representation and inserts necessary communication collectives. FastFold also introduced Duality Async Operations (DAO) for Pytorch in order to overlap the communication with computation, and AutoChunk to optimize chunk range and size.  While DAP serves as an early conceptual precursor to CP, it was constrained to the Evoformer architecture in AF2. 

At the system level, ScaleFold \cite{scalefold} combined Dynamic Axial Parallelism (DAP) with compilation, CUDA Graphs, efficient low-level kernels and batched GEMMs in order to reduce compute and memory footprint. As a result, ScaleFold enabled AF2 training across 2,080 NVIDIA H100 GPUs and reduced training time from 7 days to 10 hours. 

Extending system-level optimizations to modern structure prediction models, MegaFold \cite{la2025megafoldsystemleveloptimizationsaccelerating} accelerated AF3 training through ahead-of-time caching, Triton-based kernels, and operator fusion. Megafold reported a reduction in training peak memory usage of up to 1.23$\times$ and achieved 1.35$\times$ longer sequence lengths during training. 

In addition to these computational methods, there are other strategies that mitigate memory limits through architectural approximations. For example, SeedFold \cite{seedfold} reduces computational and memory costs by replacing standard triangular attention with a linear triangular attention mechanism. Once linearized forms of attention are adopted, sequence-parallel methods such as LASP \cite{lasp} and LASP-2 \cite{lasp2} enable extreme context lengths as shown for LLMs, but linearized attention does not preserve the dense pairwise interaction structure used in AlphaFold-style biomolecular modeling.

The Fold-CP framework diverges from the previous work: It directly tackles the memory wall of modern AF3-like architectures without chunking, batching, or attention approximations. Whereas DAP shards the pair tensor along a single axis, Fold-CP tiles it across a full $\sqrt{P}\times\sqrt{P}$ device grid, achieving $O(N^2/P)$ memory per rank rather than $O(N^2/\!\sqrt{P})$ (Section~\ref{sec:sharding})
Beyond the per-GPU memory saving, Fold-CP introduces algorithmic novelty to achieve practical execution speed:  Fold-CP implements several novel algorithms that are performant on the 2D-mesh, and adapts prior 1D solutions to 2D where needed, for the critical Triangle Attention, Triangle Multiplication, Outer Product Mean, Attention with Pair Bias modules (Section~\ref{sec:algorithmic}). With these advances, Fold-CP mitigates the communication overhead and enables performant execution at scale.

\section{Fold-CP Architecture: Multi-dimensional Sharding}\label{sec:sharding}

The challenge of scaling AF3-like architectures is distinct from scaling LLMs due to the existence of different bottlenecks: unlike LLMs that process 1D sequences, a structure prediction model is a multimodal system that must synchronize three distinct geometric manifolds: the evolutionary manifold (MSA), the sequence manifold (primary structure tokens), and the Euclidean manifold (3D coordinates). The Fold-CP framework does not treat these as isolated data streams but rather as distributed shards of a single high-dimensional state that must remain coherent across a mesh of GPUs. 

Standard machine learning frameworks, such as JAX \cite{jax2018github} and PyTorch \cite{pytorch2019}, provide high-level APIs for modeling hardware grids (e.g., \texttt{DeviceMesh}) and executing cooperative operations (e.g., PyTorch \texttt{DTensor} operations). However, these high-level APIs are typically constrained to preset communication patterns, focusing on standard collective communications such as \texttt{all-gather} or \texttt{all-to-all}.
Because synchronizing the aforementioned biological manifolds requires highly irregular routing, these standard collectives are insufficient. To implement custom communication patterns, especially those that require sufficient local computation to mask communication latency, developers need to resort to lower-level APIs for peer-to-peer (P2P) \texttt{send/recv} primitives. 

Fold-CP addresses these complexities by introducing new distributed algorithms derived from first principles using \texttt{torch.distributed} primitives. This approach allows Fold-CP to precisely orchestrate communication alongside efficient local computation, for example, by interleaving asynchronous P2P transfers with the execution of highly optimized \texttt{cuEquivariance} kernels \cite{cuequivariance}.

\begin{figure}[h]
    \centering
    \includegraphics[width=0.95\textwidth]{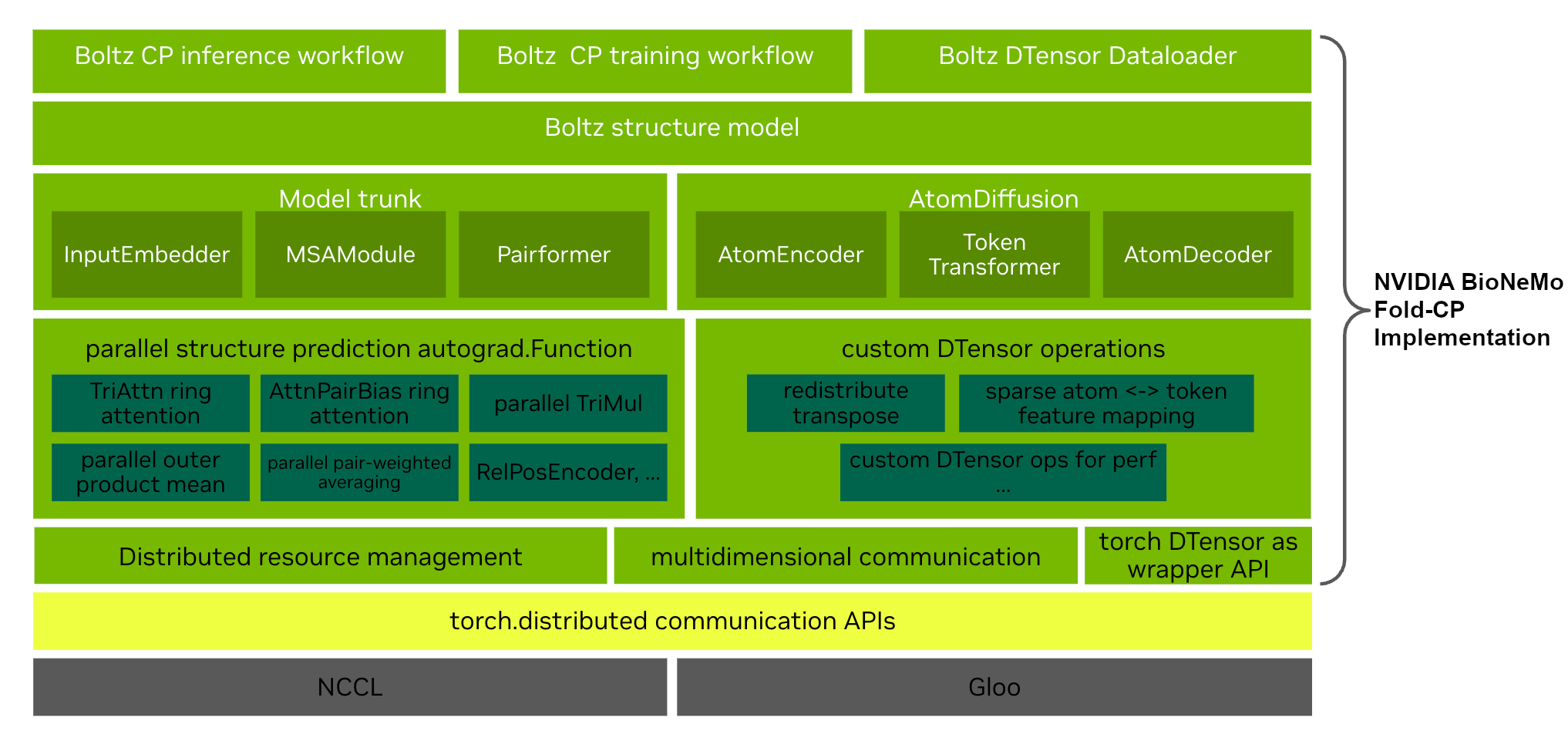} 
    \caption{\textbf{Software Architecture of the Fold-CP Framework.} Fold-CP design takes a bottom-up approach starting with foundational \texttt{torch.distributed} APIs, making module-level operators CP-aware, and incrementally building up to model-specific workflows. This design allows developers to port specific distributed modules into diverse model organizations without managing low-level synchronization logic or altering the model flow.}
    \label{fig:software_stack}
\end{figure}

Recognizing that various AF3-like models share common foundational layers yet differ in the organization of these modules, Fold-CP follows a bottom-up design
(Fig. \ref{fig:software_stack}). This architecture exposes sufficient building blocks for developers to customize the CP implementation of their models without the burden of managing the underlying distributed primitives or their performance implications. 

The remainder of this section describes the multi-dimensional activation sharding strategy covering token-pair tiling, MSA sharding, and atom sequence sharding, followed by the square topology constraint that ensures symmetric communication. Section~\ref{sec:algorithmic} then details the distributed algorithm for each computational module.

\subsection{Activation Sharding Strategy}

To achieve efficient scaling of context length with GPU count, Fold-CP implements a multi-dimensional sharding strategy covering all three geometric manifolds. This ensures that no single device holds the full global state of the biomolecule.

\subsubsection{Token Dimension and Pair Tensor Tiling}

The primary bottleneck in structure prediction is the Pair Representation tensor $z \in \mathbb{R}^{N \times N \times D}$, where $N$ is the number of input tokens representing all molecular entities including residues, nucleotides and ligand atoms and $D$ is the feature dimension. For a complex of 10,000 residues, this tensor represents $10^8$ interactions. The Fold-CP framework distributes $z$ across a grid of $P$ processors.

Unlike Data Parallelism, which replicates the model and splits the batch, Fold-CP splits the single sample. For the Pair Representation, the framework adopts a 2D tiling strategy. The global $N \times N$ matrix is partitioned into a $\sqrt{P} \times \sqrt{P}$ grid of blocks, each of size $(N/\sqrt{P}) \times (N/\sqrt{P})$. Each GPU $(r, c)$ in the device mesh manages a specific sub-block of interactions corresponding to residues in range $[r \cdot \frac{N}{\sqrt{P}}, (r+1) \cdot \frac{N}{\sqrt{P}}]$ interacting with residues in range $[c \cdot \frac{N}{\sqrt{P}}, (c+1) \cdot \frac{N}{\sqrt{P}}]$. This tiling is critical because it localizes the memory footprint of the $O(N^2)$ tensor to $O(N^2/P)$ per device.

\subsubsection{MSA Dimension Sharding}

The Multiple Sequence Alignment (MSA) representation encodes the evolutionary history of the protein. In models like Boltz-1, the size of the MSA cluster can reach 4,000 sequences to capture deep evolutionary covariation. Storing the full MSA representation ($S \times N \times D$) can be prohibitive. The framework shards this tensor along both the sequence depth ($S$) and token ($N$) dimensions, ensuring it aligns with the Pair Representation sharding. This alignment is crucial for the "Outer Product Mean" operation, which projects information from the MSA to the Pair representation. By keeping both the sequence and token dimensions sharded consistently, the projection can occur locally or with minimal communication overhead.

\subsubsection{Atom Dimension Sharding}

The Diffusion Module operates on fine-grained atom coordinates rather than coarse-grained residues. A single token (residue) maps to multiple atoms (e.g., approximately 40 atoms for a nucleic acid token comprising a sugar, phosphate, and base). The mapping is irregular, meaning a naive token-based shard would result in load imbalance, e.g. a GPU holding a glycine-rich region would have far fewer atoms than a GPU holding a tryptophan-rich region. To facilitate the frequently-used mapping between atom and token feature space throughout the model workflow, Fold-CP introduces the concept of "co-sharding" token and atom features where the shard of tokens are guaranteed to be mapped to the corresponding shard of atoms locally on the same device and vice versa. On the other hand, the distributed window batching algorithm (Section~\ref{sec:winbatch}) requires contiguous atom sequences within each shard to avoid disrupting the atom sequence local attention windows. Fold-CP includes various utility functions to transition from the atom sequence co-sharded with token to a compact contiguous sequence to work with distributed window batching.

\subsection{The Square Topology Constraint}

A critical architectural finding in the development of this framework is the requirement for a square CP world size (e.g., $CP=4$ implies a $2 \times 2$ grid; $CP=64$ implies $8 \times 8$ etc.). This design choice facilitates the symmetric sharding of the pair representations, significantly simplifying the implementation without compromising computational performance. 
While an alternative implementation supporting a generic CP device grid is possible, it would introduce a significant performance and maintainability burden, due to the axial nature of the geometric operations: the Triangle Attention and Triangle Multiplication modules enforce physical consistency through interleaved "Row-wise" and "Column-wise" updates. In the Row-wise update, information propagates among all residues interacting with a target residue $i$ (the $i$-th row of the pair matrix). In the Column-wise update, information propagates among all residues that target residue $j$ interacts with (the $j$-th column). In a distributed setting, a GPU owning block $(r, c)$ holds a partial view of the rows and columns. To perform a Row-wise update efficiently, the device mesh typically treats the row-group of GPUs as a communicative unit. To switch to a Column-wise update, the communication pattern must transpose. If the device mesh were rectangular (e.g., $2 \times 8$), transposing the data layout would require complex re-indexing and load balancing, as the number of devices in a row-group would differ from the number in a column-group. By imposing a square topology ($R = C = \sqrt{P}$), the framework ensures that the communication volume and latency for row and column operations are identical. This symmetry simplifies distributed ring communication algorithms by means of circulant shift of the data buffer among the involved GPUs~\cite{ringcomm}. 

\section{Fold-CP Architecture: Distributed Operations}\label{sec:algorithmic}

The implementation of Fold-CP for AF3-like models extends beyond standard tensor sharding; it requires the redesign of core algorithmic primitives to function over a distributed mesh without materializing the global state. This section details the distributed algorithm for each computational module. Table~\ref{tab:cp_complexity} provides a complexity overview; the following subsections describe the algorithms in depth.

\subsection{The Custom Autograd Imperative}\label{sec:autograd}

A foundational design decision, visible at the \texttt{autograd.Function} layer in the software stack (Figure~\ref{fig:software_stack}), underlies every distributed module described below. PyTorch's native Distributed Tensor (DTensor) operations frequently trigger implicit All-Gather behavior in the backward pass. When backpropagating through a sharded operation, DTensor may attempt to gather the full global inputs to compute gradients, inadvertently reconstructing the $O(N^2)$ tensor on a single device and causing immediate Out-Of-Memory (OOM) failures.

To prevent this, in its current form, the Fold-CP framework adopts a strict principle: no distributed module may rely on native DTensor operator dispatch for differentiable paths. Instead, each module implements a custom \texttt{torch.autograd.Function} that explicitly defines both the forward and backward passes using the same distributed ring algorithm, ensuring that gradients are scattered and reduced stage-wise without ever materializing the global tensor. This design decision applies universally to every module described in Sections~\ref{sec:triattn}--\ref{sec:cueq} and is a prerequisite for correctness at scale. Modifying the native DTensor behavior to be customizable in a way to avoid unnecessary collective communication primitives and performance pitfalls is left for future work.

\begin{table}[t]
\centering
\footnotesize
\begin{tabular}{@{}llllll@{}}
\toprule
\textbf{Component} & \textbf{Serial Time} & \textbf{Serial Space} & \makecell{\textbf{CP Time}\\ (per rank)} & \makecell{\textbf{CP Space}\\ (per rank)} & \textbf{Comm.\ Pattern} \\ \midrule
Triangle Attention 
& $BHN^3D$  & $BHN^2D$  & $BHN^3D/P$  & $BHN^2D/P$  & 2D ring + transpose \\

Triangle Multiplication 
& $BN^3D$ 
& $BN^2D$ 
& $BN^3D/P$ 
& $BN^2D/P$ 
& 2D ring (Cannon-style) \\

Pair Weighted Avg. 
& $BHSN^2D$ 
& $BHSND$ 
& $BHSN^2D/P$ 
& $BHSND/P$ 
& 2D ring + tiled softmax \\

Outer Product Mean 
& $BSN^2D^2$ 
& $BN^2D^2$ 
& $BSN^2D^2/P$ 
& $BN^2D^2/P$ 
& 2D ring + all-reduce \\

Attn Pair Bias (Ring) 
& $BHN^2D$ 
& $BHND$ 
& $BHN^2D/\sqrt{P}$ 
& $BH\frac{N}{\sqrt{P}}D$ 
& 1D ring + transpose \\

Attn Pair Bias (Shard) 
& $BAHD$ 
& $BAHD$ 
& $B\frac{A}{\sqrt{P}}HD$ 
& $B\frac{A}{\sqrt{P}}HD$ 
& None (upstream gather) \\

Window Batching 
& $BA^2D$ 
& $A^2 + BAD$ 
& $B\frac{A}{\sqrt{P}}D$ 
& $B\frac{A}{\sqrt{P}}D$ 
& P2P halo exchange \\
\bottomrule
\end{tabular}
\caption{\textbf{Complexity of distributed modules and the scaling factors.} $N$: number of tokens; $P$: GPU count; $B$: batch size; $S$: number of MSA sequences; $H$: number of attention heads; $D$: feature dimension (per-head dimension for attention-based modules); $A$: number of atoms. Shard-wise complexity absorbs constant window and key-span sizes into $O(\cdot)$; atoms are partitioned into fixed-size windows with overlapping key neighborhoods. All ring patterns use $\sqrt{P}$ steps with double-buffered overlap. For Window Batching, the serial column reflects the pre-Toeplitz baseline; the CP column includes both the Toeplitz optimization and distribution (see Section~\ref{sec:winbatch}).}
\label{tab:cp_complexity}
\end{table}

\begin{figure}[!ht]
    \centering
    \includegraphics[width=\textwidth]{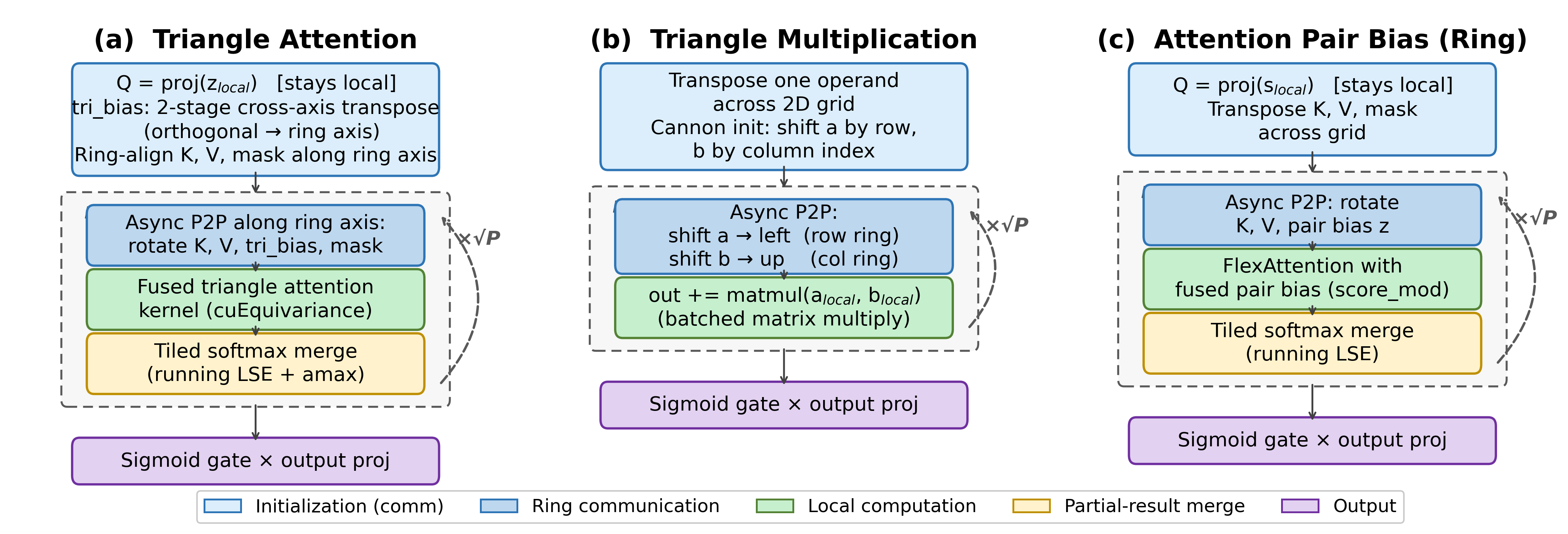}
    \caption{\textbf{Distributed forward-pass flow for the three core ring-communication modules.}
    \textbf{(a)}~Triangle Attention: cross-axis transpose of triangular bias followed by a 1D ring rotation of keys, values, and bias with tiled-softmax merge.
    \textbf{(b)}~Triangle Multiplication: Cannon-style 2D ring where projections shift along orthogonal axes with additive accumulation.
    \textbf{(c)}~Ring Attention Pair Bias: 1D ring rotation of keys, values, and pair bias with tiled-softmax merge.
    Pair Weighted Averaging (Section~\ref{sec:pwa}) shares the softmax-merge archetype of~(a)/(c) over a 2D ring;
    Outer Product Mean (Section~\ref{sec:opm}) shares the Cannon accumulation archetype of~(b), replacing matrix multiplication with outer products.}
    \label{fig:cp_flow}
\end{figure}

\subsection{Triangle Attention}\label{sec:triattn}

Triangle Attention updates the pair representation by performing multi-head attention over the rows (or columns) of the pair tensor, with an additive triangular bias derived from the same representation:
\begin{equation}
Q, K, V = \mathrm{proj}(z), \quad \mathrm{output} = \mathrm{softmax}\!\left(\frac{QK^T}{\sqrt{d}} + \mathrm{tri\_bias}(z) + \mathrm{mask\_bias}\right) V
\end{equation}
where $z \in \mathbb{R}^{B \times N \times N \times D}$ is the pair representation. The triangular bias operates along an axis orthogonal to the padding mask, requiring the algorithm to process information from all three axes $\{i, j, k\}$ despite the pair representation being sharded on only two.

\textbf{Sharding and algorithm.}
The pair representation is tiled as $(N/\sqrt{P}, N/\sqrt{P})$ blocks on the 2D device mesh. Queries remain local on each rank. Keys, values, triangular bias, and the padding mask are redistributed through a two-phase process: an initial transpose-based redistribution aligns the bias data across the grid, followed by a ring of $\sqrt{P}$ steps where each rank computes partial attention scores on the arriving chunk and merges partial outputs using numerically stable tiled softmax with running log-sum-exp accumulators. This preserves exact equivalence to serial softmax without materializing the full $N \times N$ attention matrix on any rank. Double buffering overlaps the P2P transfer of the next chunk with the current step's computation (Fig.~\ref{fig:cp_flow}a).

\textbf{Complexity and scaling.}
Serial compute is $O(BHN^3D)$. Under Fold-CP, each rank performs $O(BHN^3D/P)$ compute and stores $O(BH(N/\sqrt{P})^2D)$ activations. Each ring step transfers $O(BH(N/\sqrt{P})^2D)$ data. The ratio of compute to communication per step scales as $O(N)$, so as the sequence length grows, the algorithm becomes increasingly communication-efficient.

Triangle Attention integrates with cuEquivariance fused kernels for additional acceleration (Section~\ref{sec:cueq}).

\subsection{Triangle Multiplication}\label{sec:trimul}

Triangle Multiplication aggregates over one index of the pair representation to update the other two. The outgoing and incoming variants are:
\begin{align}
\mathrm{Outgoing:} \quad z_{nm} &\mathrel{+}= \sum_k a_{nk} \odot b_{mk} \\
\mathrm{Incoming:} \quad z_{nm} &\mathrel{+}= \sum_k a_{kn} \odot b_{km}
\end{align}
where $a, b$ are projections of the pair representation $z \in \mathbb{R}^{B \times N \times N \times D}$. This operation is structurally a batched matrix multiplication with element-wise gating, inspired by the geometric consistency arising from triangle inequality.

\textbf{Sharding and algorithm.}
The pair representation is tiled as $(N/\sqrt{P})^2$ blocks. One operand is transposed across the 2D grid to align the contracted index with the ring shift direction. The algorithm then proceeds in a Cannon-style ring \cite{cannon}: in each of $\sqrt{P}$ steps, ranks perform a local batched matrix multiplication between resident blocks of $a$ and $b$ while simultaneously initiating asynchronous P2P transfers to circulate $a$ along rows and $b$ along columns. Partial results accumulate into the output block. The outgoing and incoming variants differ only in which operand is transposed, reusing the same ring communication pattern (Fig.~\ref{fig:cp_flow}b).

\textbf{Complexity and scaling.}
Serial compute is $O(BN^3D)$. Under Fold-CP, each rank performs $O(BN^3D/P)$ compute and stores $O(B(N/\sqrt{P})^2D)$ activations. Each ring step transfers
$O(B(N/\sqrt{P})^2D)$. The ratio of $O(N^3/P)$ compute to $O(N^2/\sqrt{P})$ communication grows linearly with token count, making the algorithm increasingly efficient in masking communication for larger biomolecular complexes.

\subsection{Pair Weighted Averaging}\label{sec:pwa}

Pair Weighted Averaging uses the pair representation to form attention weights over the sequence dimension, then applies them to a value representation projected from the MSA:
\begin{equation}
w = \mathrm{softmax}(\mathrm{proj}_z(z) + \mathrm{mask\_bias}), \quad o_{si} = \sum_j w_{ij} \cdot v_{sj}, \quad \mathrm{output} = \mathrm{proj}_o(g \cdot o)
\end{equation}
where $v, g$ are projected from the MSA representation $m \in \mathbb{R}^{B \times S \times N \times D}$, $z \in \mathbb{R}^{B \times N \times N \times D}$ is the pair representation, and $o \in \mathbb{R}^{B \times S \times N \times D}$ is the weighted output. The softmax and weighted sum over the $N^2$ pair dimension require distribution for large $N$.

\textbf{Sharding and algorithm.}
The MSA representation is sharded along both the sequence depth $S$ and token dimension $N$, consistent with the pair sharding. Pair weights are transposed across the 2D grid, then ring-shifted. Each ring step computes softmax on a local weight block and an einsum with the corresponding value block; partial outputs are merged using tiled softmax with running amax and log-sum-exp accumulators, ensuring numerical equivalence to the global softmax without ever materializing the full $N \times N$ weight matrix on any rank. This communication pattern parallels the softmax-merge ring of Triangle Attention and Attention Pair Bias (Fig.~\ref{fig:cp_flow}a,c), extended to a 2D ring.

\textbf{Complexity and scaling.}
Serial compute is $O(BHSN^2D)$. Under Fold-CP, each rank performs $O(BHSN^2D/P)$ compute and stores $O(BH(S/\sqrt{P})(N/\sqrt{P})D)$ activations. Each ring step transfers $O(BH(S/\sqrt{P})(N/\sqrt{P})D)$. As with the previous modules,  arithmetic intensity scales as $O(N)$, improving linearly with token count.

\subsection{Outer Product Mean}\label{sec:opm}

Outer Product Mean builds a pair representation from the MSA representation by averaging outer products over the sequence dimension:
\begin{equation}
z_{ijcd} = \frac{1}{|S|} \sum_s a_{sic}\, b_{sjd}, \quad a = \mathrm{proj}_a(m),\; b = \mathrm{proj}_b(m)
\end{equation}
where $a, b \in \mathbb{R}^{B \times S \times N \times D}$ are projections of the MSA representation $m$. The indices $c, d$ each range over $D$, so the outer product accumulates a $D \times D$ matrix per pair $(i,j)$, yielding an intermediate $z \in \mathbb{R}^{B \times N \times N \times D \times D}$ that is flattened and linearly projected to $\mathbb{R}^{B \times N \times N \times D}$. The resulting $O(S \times N^2 \times D^2)$ compute and $O(N^2 \times D^2)$ memory require distribution for large $N$ and $S$.

\textbf{Sharding and algorithm.}
The MSA is sharded along $S$ and $N$. Projected tensor $a$ is transposed across the 2D grid for ring alignment; $b$ is initialized along columns. In each of $\sqrt{P}$ ring steps, ranks accumulate partial outer products by shifting $a$ along rows and $b$ along columns. A single all-reduce finalizes the mask count for the mean. The backward pass uses two independent ring loops for $\nabla a$ and $\nabla b$. This Cannon-style 2D ring with additive accumulation mirrors the Triangle Multiplication pattern (Fig.~\ref{fig:cp_flow}b), with outer products replacing matrix multiplication and a final all-reduce for the mean normalization.

\textbf{Complexity and scaling.}
Serial compute is $O(BSN^2D^2)$. Under Fold-CP, each rank performs $O(BSN^2D^2/P)$ compute and stores $O(B(N/\sqrt{P})^2D^2)$ activations. Each ring step transfers $O(B(S/\sqrt{P})(N/\sqrt{P})D)$. The arithmetic intensity scales as $O(N \cdot D)$, making this module particularly communication-efficient for large systems.

\subsection{Attention Pair Bias}\label{sec:apb}

Attention Pair Bias implements standard multi-head attention with an additive pair bias $z$ on the attention logits:
\begin{equation}
\mathrm{attn} = \mathrm{softmax}\!\left(\frac{QK^T}{\sqrt{d}} + z + \mathrm{mask\_bias}\right), \quad \mathrm{output} = \mathrm{proj}_o(g \cdot (\mathrm{attn} \cdot V))
\end{equation}
where $Q, K, V, g$ are projected from the single representation $s$. This operation appears in two contexts that require different distribution strategies.

\textbf{Ring variant} (full-sequence attention in the Pairformer and token transformer). This variant handles the dense $N \times N$ interactions sharded across the 2D device mesh. Queries remain local while keys, values, pair bias, and mask are transposed for ring alignment and then rotated through $\sqrt{P}$ ring steps. At each step, ranks compute partial attention scores and merge them with tiled softmax, maintaining exact numerical equivalence to the serial implementation. (Fig.~\ref{fig:cp_flow}c).

\textbf{Shard-wise variant} (window-batched atom attention). This variance handles attention computed over local windows of atoms and it is designed to be communication-isolated: Each rank holds $K/\sqrt{P}$ windows of queries, keys, values, and pair bias and computes standard multi-head attention locally over its windows. The correctness relies on upstream window batching and distributed gather operations (Section~\ref{sec:winbatch}) supplying each rank with exactly the data needed. This combination of window-based sparsity ($O(W \cdot W_k)$ per window, where $W_k$ is the key neighborhood span, with AF3's default constants $W=32$ and $W_{k}=128$ instead of $O(A^2)$ where $A$ is atom count) with Fold-CP ($K/\sqrt{P}$ windows per rank) allows atom-level attention to scale without ring communication.

\textbf{Complexity and scaling.}
Ring variant: serial $O(BHN^2D)$, Fold-CP $O(BHN^2D/\sqrt{P})$ compute per rank, $O(BH(N/\sqrt{P})D)$ space; arithmetic intensity scales as $O(N)$. Shardwise variant: $O(B(A/\sqrt{P})HD)$ per rank with no communication overhead.

Both variants leverage FlexAttention kernel fusion for pair-bias addition (Section~\ref{sec:cueq}).

\subsection{Window Batching and Distributed Gather}\label{sec:winbatch}

Atom-level attention in structure prediction models requires each atom to attend to its spatial neighbors, but atoms that are neighbors in 3D space are generally non-contiguous in sequence memory. For complexes with $A > 100{,}000$ atoms, naive self-attention has $O(A^2)$ cost. Serial window batching groups spatially proximate atoms into fixed-size windows $W$, reducing attention to $O(A \cdot W)$, but the original Boltz implementation materializes a $(K, K)$ sparse indexing matrix (where $K = A/W$ is the number of windows) and applies it via dense einsum over the feature dimension, yielding $O(BA^2D)$ time and $O(A^2 + BAD)$ space, $O(A^2)$ for the indexing matrix itself, plus $O(BAD)$ for the input and output atom feature tensors, for the window construction alone, before any attention is computed. In the following we list approaches used in Fold-CP to scale this operation:

Careful examination revealed that the window batching indexing matrix has a block-Toeplitz structure: constant along diagonals with a $+2$ shift between consecutive windows. This mathematical property enables replacing the sparse matrix with an $O(1)$-memory virtual view, achieving the same $O(A \cdot W_k)$ computation with zero matrix allocation and superior cache locality.
The Toeplitz structure exhibits translational symmetry: each GPU can compute its local windows using coordinate-adjusted offsets, requiring only small halo exchanges ($\sim$48 atoms) with neighboring ranks. This achieves linear scaling across GPUs with no global synchronization.

The token-to-atom representation mapping traditionally requires a dense matrix multiplication over all tokens. Fold-CP introduces distributed gather operations that exploit the fact that atoms within a window, map to contiguous token ranges. Using bounding-box interval-based P2P communication, each rank fetches only the token representations overlapping its local atom range, with a communication volume of $O(\delta \cdot D)$ per window where $\delta$ is the local token interval, typically 5-10 tokens.

These approaches form the upstream pipeline that enables communication-free shard-wise attention (Section~\ref{sec:apb}). The data pipeline pre-computes window assignments and distributes them across ranks, so the attention module receives self-contained local shards requiring no further inter-GPU communication.

\textbf{Complexity and scaling.}
Each rank constructs $O(A/\sqrt{P})$ local windows via the Toeplitz virtual view, exchanging only a small constant-size halo with neighboring ranks at shard boundaries. Per-rank compute and space are both $O(B(A/\sqrt{P})D)$. As an example, this design reduced ribosome-scale inference ($\sim$150k atoms) from 3 hours on 576 devices to 1 hour on 64 devices, resulting in a 27$\times$ improvement in GPU-hour efficiency.

\subsection{cuEquivariance and Kernel Fusion}\label{sec:cueq}

To maximize throughput on NVIDIA GPU architectures, the framework integrates two kernel fusion strategies. For Triangle Attention, fused kernels in cuEquivariance \cite{cuequivariance} are used for acceleration. cuEquivariance supports BF16 precision, reducing memory footprint by 50\% compared to FP32 without compromising numerical stability. This kernel fusion was the primary enabler for reaching token counts of $\sim$10,500 on 144 NVIDIA H100 GPUs .

For Attention Pair Bias (both ring and shard-wise variants), FlexAttention's \texttt{score\_mod} mechanism fuses the pair-bias addition into the attention kernel, avoiding materialization of the full $N \times N \times H$ bias tensor. These kernel optimizations are complementary to distributed algorithms: each ring step or local window computation calls the fused kernel on its local data block, combining algorithmic and hardware-level efficiency.

\section{Results}

\subsection{Scaling Benchmarks for Inference and Training Context Length}\label{sec:inference_training_scaling}
\begin{figure}[!ht]
    \centering
    \includegraphics[width=1.0\textwidth]{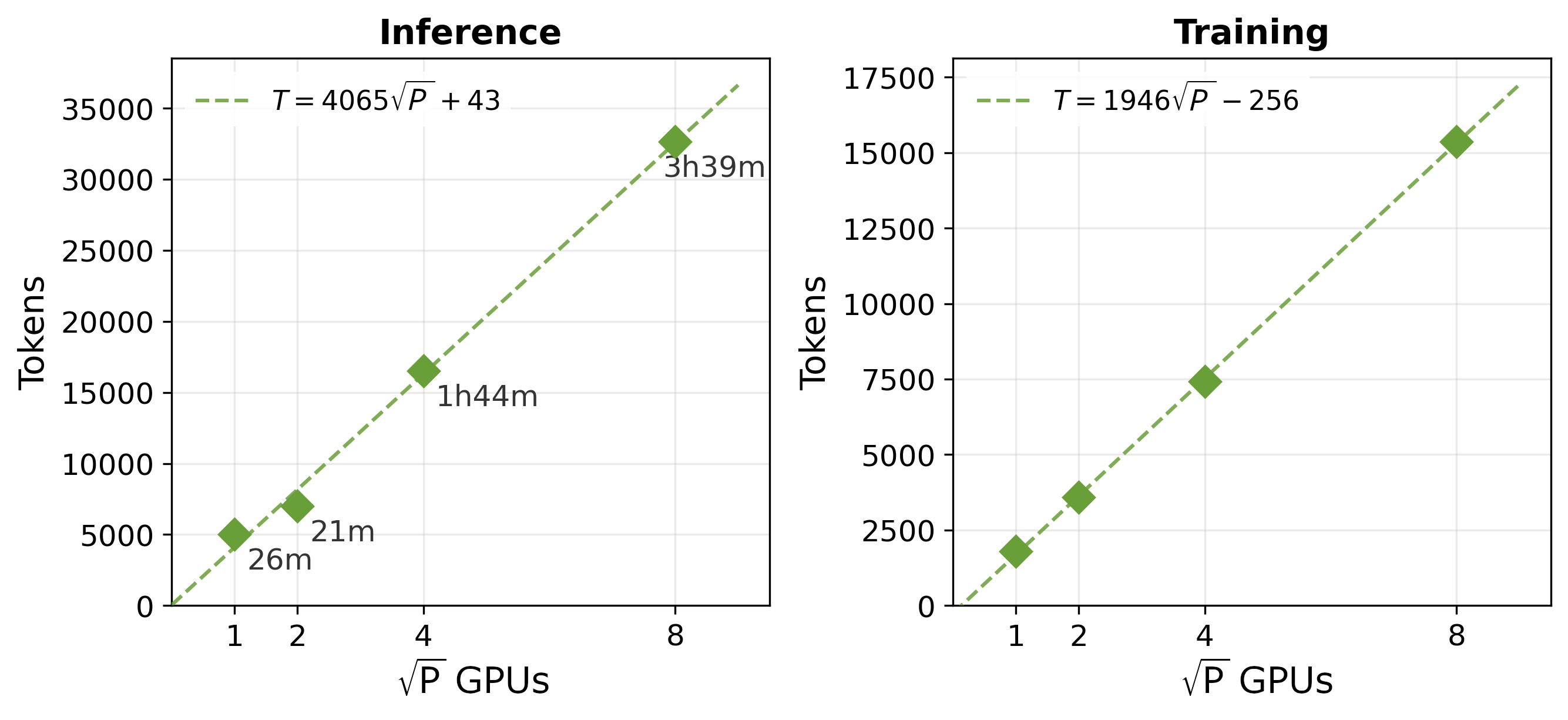}
    \caption{\textbf{Boltz-2 CP scalability on NVIDIA B300 GPUs via maximum context length.} For a given GPU count $P$, maximum context length reached is plotted. Linear trend lines are used to inform  with respect to number of GPUs in a CP rank, $\sqrt{P}$, due to the square topology requirement of Fold-CP. (Left) Inference scaling performance: The maximum context length scales linearly at a rate of approximately 4,000 tokens/$\sqrt{P}$, with runtime annotated accordingly. (Right) Training scaling performance: The training follows a similar scaling with a approximately half the slope of that of inference.}
    \label{fig:boltz2_inference_scaling}
\end{figure}

The Fold-CP framework enables the extension of maximum context length for both inference and training. In inference benchmarks, Boltz-2 with Fold-CP, i.e. Boltz-2 CP, achieves linear scaling with the square root of the number of GPUs ($\sqrt{P}$). Measured on up to 64 NVIDIA B300 GPUs, using BF16 precision, structure prediction of biomolecules of up to 32k tokens can be achieved without chunking (Figure \ref{fig:boltz2_inference_scaling}). This scaling trend remains consistent during training, albeit with a reduced slope of approximately 1,900 tokens/$\sqrt{P}$, compared to approximately 4,000 tokens/$\sqrt{P}$ observed during inference.

We further validate the scaling of the Fold-CP framework by repeating the benchmarks with Boltz-1, i.e. Boltz-1 CP, on NVIDIA H100 GPUs using FP32 precision, where we observed a consistent linear scaling trend (Table \ref{tab:boltz1_inference_scaling}). Notably, the inclusion of confidence prediction results in roughly 10\% reduction in maximum context length due to the associated memory overhead.  

\begin{table}[!ht]
    \centering
    \footnotesize
    \begin{tabular}{@{}ccccc@{}}
        \toprule
        \multirow{2}{*}{\textbf{\# GPUs}} & 
        \multicolumn{2}{c}{\textbf{Confidence prediction off}} & 
        \multicolumn{2}{c}{\textbf{Confidence prediction on}} \\
        \cmidrule(lr){2-3} \cmidrule(lr){4-5}
        & \textbf{Tokens} & \textbf{Walltime (mins)} & \textbf{Tokens} & \textbf{Walltime (mins)} \\
        \midrule
        $1\times1$ & 2,700 & 13 & 2,400 & 12 \\
        $2\times2$ & 4,000 & 13 & 3,800 & 16 \\
        $4\times4$ & 6,500 & 37 & 6,000 & 33 \\
        $8\times8$ & 11,500 & 47 & 11,000 & 55 \\
        \bottomrule
    \end{tabular}
    \caption{\textbf{Boltz-1 CP inference scaling on NVIDIA H100 80GB GPUs.} Confidence prediction requires caching activation from structure prediction, leading to about 10\% reduction in maximum context length. The reported token count is the output of tokenization process, e.g.\ a sample with PDB ID: 7NYC consists of 6,279 residues and gets tokenized to 6,500 tokens.}
    \label{tab:boltz1_inference_scaling}
\end{table} 

\subsection{Accuracy Benchmarks for Inference and Training }\label{sec:accuracy_consistency}
\begin{figure}[!ht]
  \centering
  \includegraphics[width=0.5\textwidth]{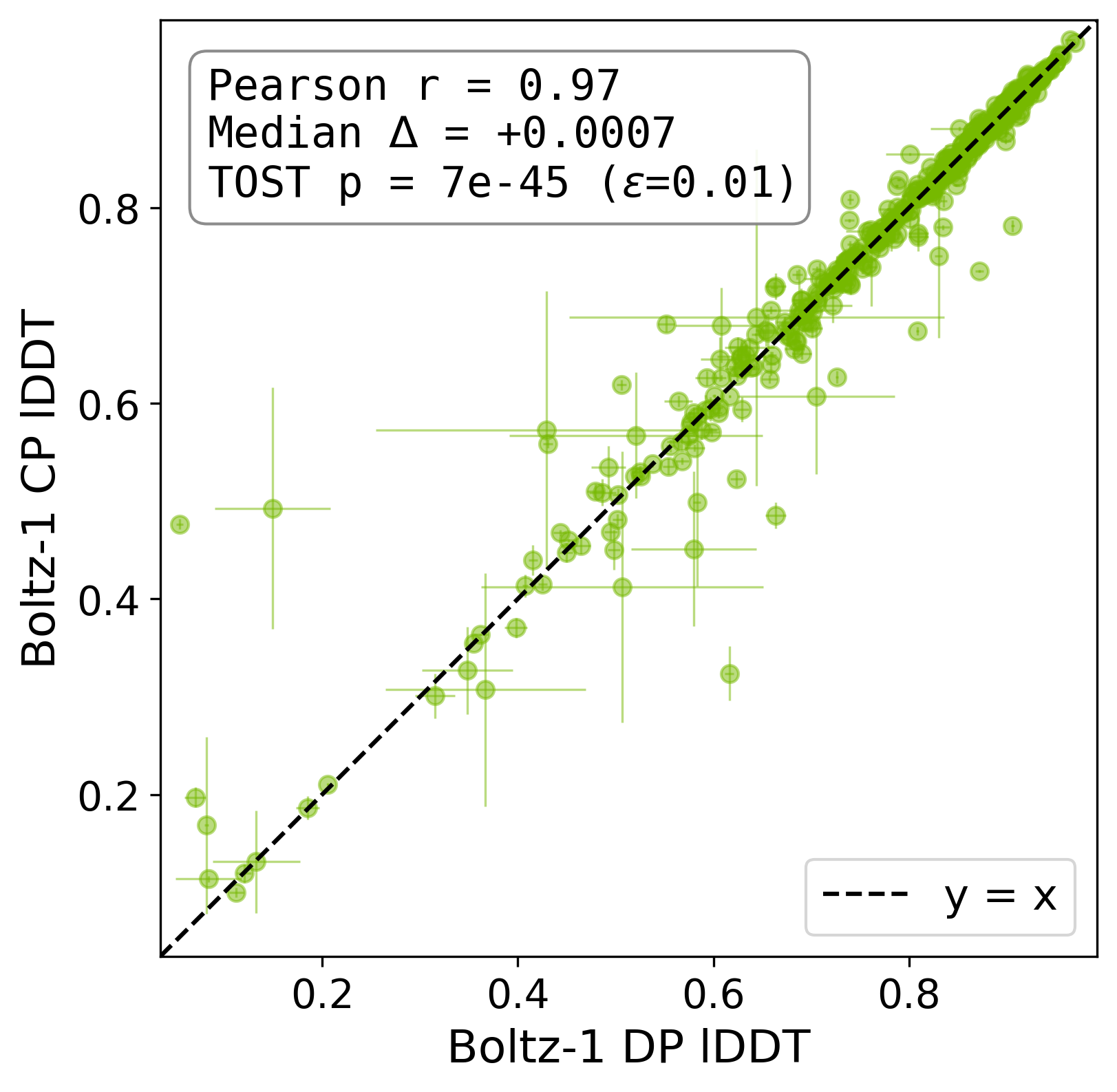}
  \caption{\textbf{Inference consistency validation on Boltz-1 test set.} Numerical equivalence between the DP baseline and Fold-CP implementation verified via a Two One-Sided Tests (TOST) procedure on a Wilcoxon signed-rank test with a margin of error ($\epsilon$=0.01). Error bar estimated from standard error of mean (SEM) from 5 diffusion samples for the respective x- (DP, horizontal bar) and y-axis (CP, vertical bar).}
  \label{fig:boltz1_inference_consistency}
\end{figure}

To establish the numerical and convergence integrity of the Fold-CP framework, we evaluate inference consistency on the Boltz-1 test set with and without Fold-CP, i.e. in the case of data parallelism alone (DP), using the Boltz-1 model. We observe a high correlation between predictions from the original DP implementation and Fold-CP framework, achieving a Pearson’s $R = 0.97$ with a median lDDT difference of $0.0007$ (Figure \ref{fig:boltz1_inference_consistency}). During these benchmarks, inputs and activations are sharded across 4 GPUs, in the aforementioned square topology ($2 \times 2$).

\begin{figure}[!ht]
  \centering
  \includegraphics[width=0.55\textwidth]{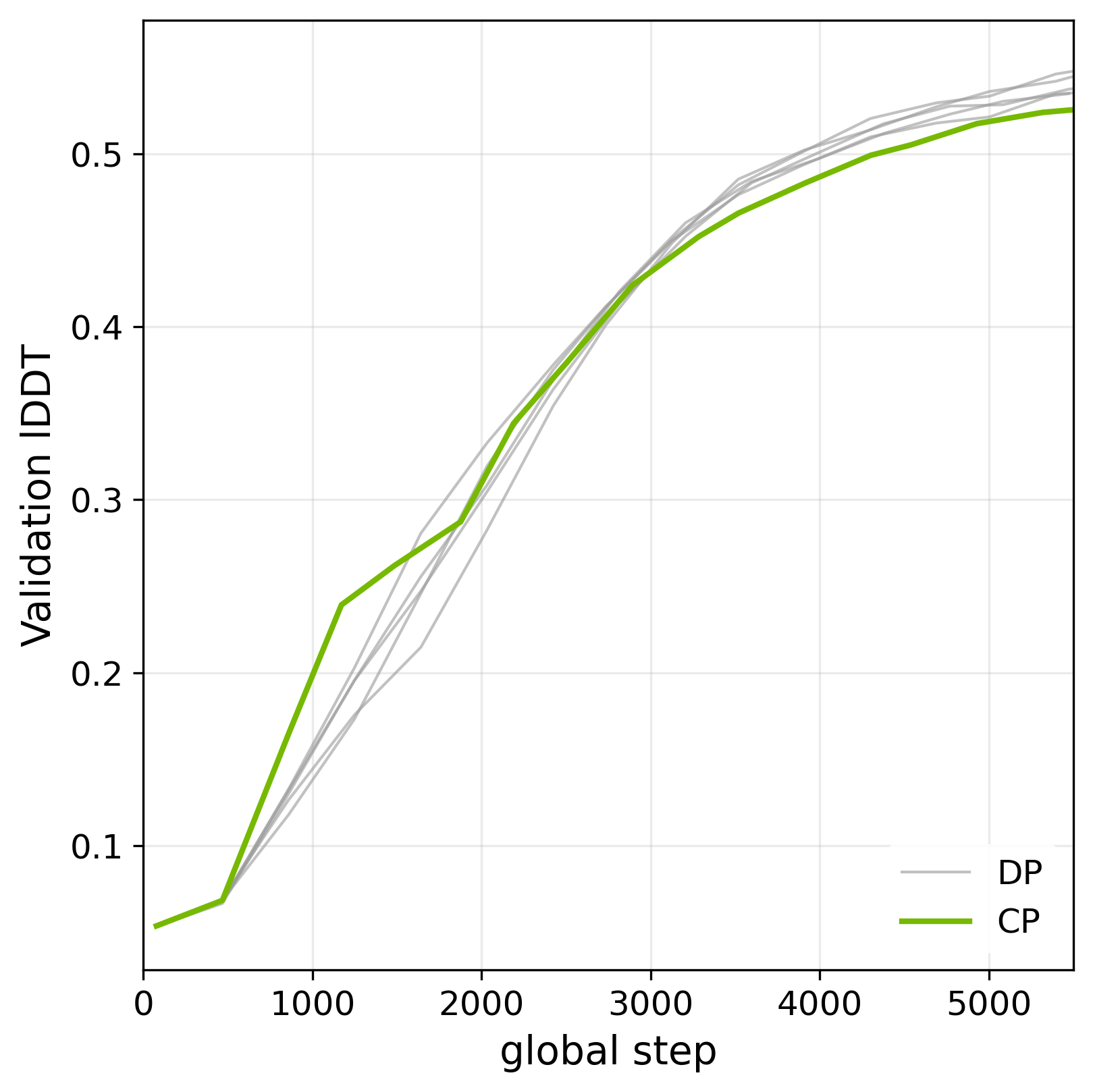}
  \caption{\textbf{Training consistency validation.} Validation lDDT trajectories for five DP baseline with random seeding and one $\text{CP}=2\times2$ configuration with a model trained on 256-token crop size. The Fold-CP validation lDDT curve largely mirrors the DP references, with variation stemming from stochastic nature of the model. The model used in this validation experiment is a truncated Boltz-1 architecture, hence it is not expected to match the published reference. See text for details.}
  \label{fig:boltz1_training_consistency}
\end{figure}

We demonstrate consistent training trajectories across DP and Fold-CP configurations. To facilitate rapid validation, we utilized a truncated Boltz-1 model architecture in which the depth of the MSA, Pairformer, and structure module stacks was reduced to 3, 12, and 12 blocks respectively with a crop size of 256 tokens. The validation lDDT for the Fold-CP configuration closely mirrors the baseline over the first 5.5k steps tested (Figure \ref{fig:boltz1_training_consistency}). We repeat the DP training multiple times to emphasize the variation in training dynamics due to the stochasticity of the model.

\newpage

\section{Case Studies}

\subsection{Boltz-2: Preventing Fragmentation Bias in the PI4KA Lipid Kinase Complex}

\begin{figure}[h]
    \centering
    \includegraphics[width=0.5\textwidth]{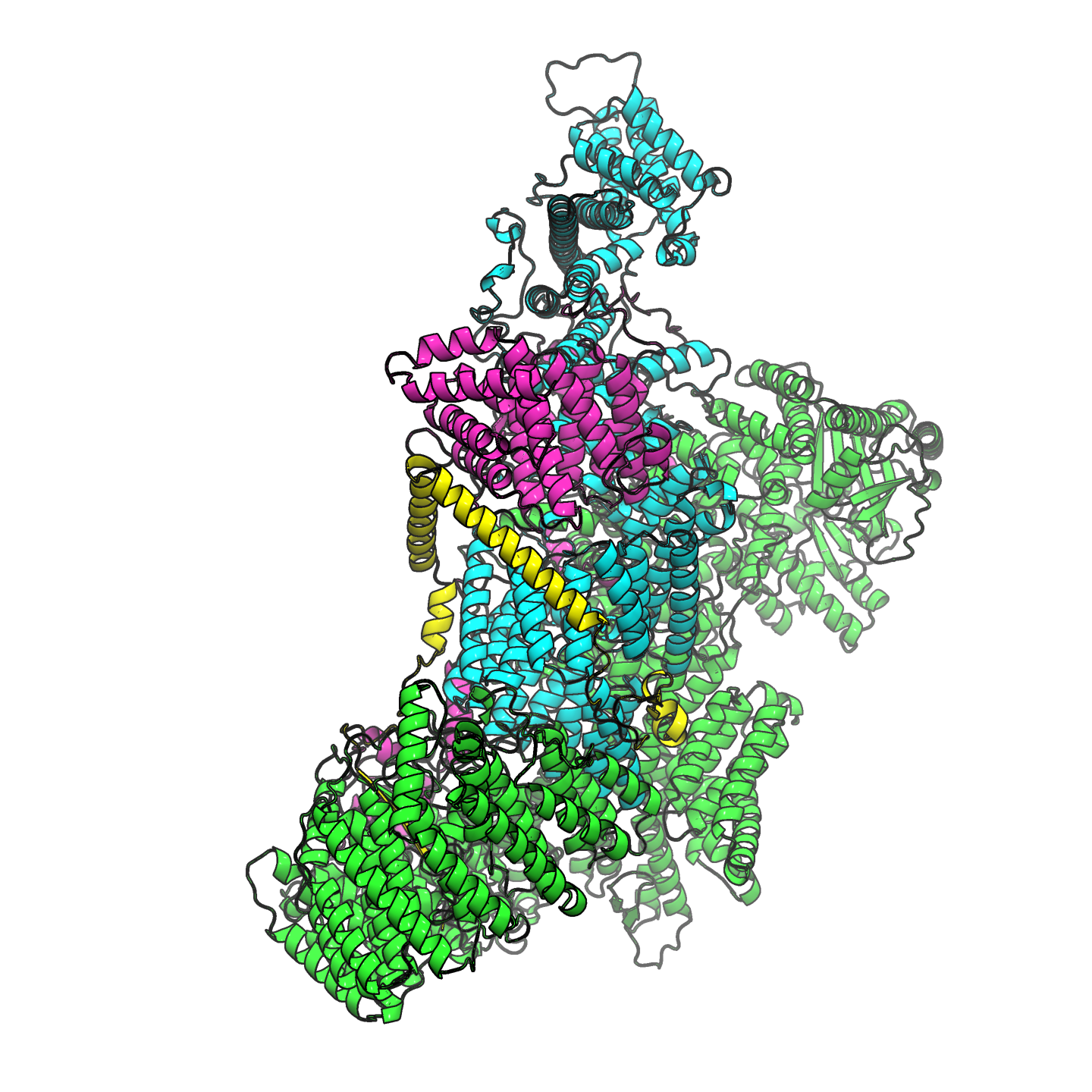}
    \caption{\textbf{Boltz-2 prediction of the TTC7A/PI4KA/FAM126A/EFR3A(700--823) complex (3,605 residues)}. PI4KA (green), TTC7A (cyan), FAM126A (magenta), and the intrinsically disordered EFR3A C-terminus (yellow). The prediction was generated in a single inference pass using Fold-CP across 4 NVIDIA H100 80GB GPUs.}
    \label{fig:Boltz-2-case-study}
\end{figure}

Multi-chain assemblies where interfaces span hundreds to thousands of residues across subunits are systematically underexplored because they exceed single-GPU context limits. The PI4KA lipid kinase complex perfectly exemplifies this gap. This multi-subunit assembly, comprised of PI4KA (lipid kinase), TTC7 (scaffold), and FAM126 (stabilizer), is recruited to the plasma membrane by EFR3, where it synthesizes signaling precursor molecules~\cite{balla2013}. Loss-of-function mutations in the TTC7A paralog cause multiple intestinal atresias and very early onset inflammatory bowel disease~\cite{avitzur2014}, yet no experimental structure exists of a complex containing TTC7A. Critically, no single intra-complex interface is sufficient to maintain a stable PI4KA complex~\cite{lees2017}; three discrete regions of PI4KA contact TTC7 across more than 1,000 sequence residues. Because this structural integrity relies on widely distributed interfaces, traditional short-context cropping pipelines invariably fail, increasing the likelihood of modeling biologically irrelevant states by exposing normally buried interfaces to the solvent.

To resolve this without fragmentation bias, we applied the Fold-CP framework to Boltz-2, modeling a heterotrimer unit of the PI4KA complex bound to the intrinsically disordered EFR3A C-terminus in a single, 3,605-residue inference across 4 NVIDIA H100 GPUs. The system distributed the attention computation to generate five structural samples in under five minutes ($\sim$54 seconds per sample) while maintaining all long-range inter-subunit contacts natively within the model's context window. Evaluated against a paralogous cryo-EM structure (PDB 9BAX~\cite{suresh2024}), the prediction achieved a TM-score of 0.83, a backbone lDDT of 0.72, and a DockQ of 0.56, which is remarkably strong given the $\sim$50\% sequence identity between the modeled TTC7A and the reference TTC7B paralogs (Figure \ref{fig:Boltz-2-case-study}). In addition to recapitulating known interactions, the uncropped inference successfully predicted a novel interaction between the EFR3A C-terminus and a large pocket at the TTC7A--PI4KA interface, revealing a topological insight that would be impossible to predict without modeling all subunits simultaneously.

\subsection{Rezo Therapeutics: Scaling AI-Guided Discovery for Massive Protein Complexes}

A key challenge in drugging protein-protein interactions (PPIs) and complexes is obtaining high-quality structural starting points for biological validation and druggability assessment. Compared to monomeric proteins, far fewer high-quality structures exist for PPIs, and the potential combinatorial possibilities of protein complexes are immense. To examine the impact that existing size constraints impose on known protein complexes, Rezo extracted all human protein complexes from the 2024 CORUM database~\cite{corum2024}.  Strikingly, less than 30\% of these CORUM complexes can be scored with the serial Boltz-1x model, meaning the majority of these complexes are inaccessible. Rezo's work with NVIDIA to combine Boltz-1 with Fold-CP (Boltz-1 CP), along with the addition of \texttt{cuEquivariance}, directly tackles this bottleneck. Rezo has now predicted high-quality structures for complexes up to 6,500 residues (encompassing 93\% of CORUM) with the potential to fold even larger complexes.

\begin{figure}[H]
    \centering
    \includegraphics[width=0.65\textwidth]{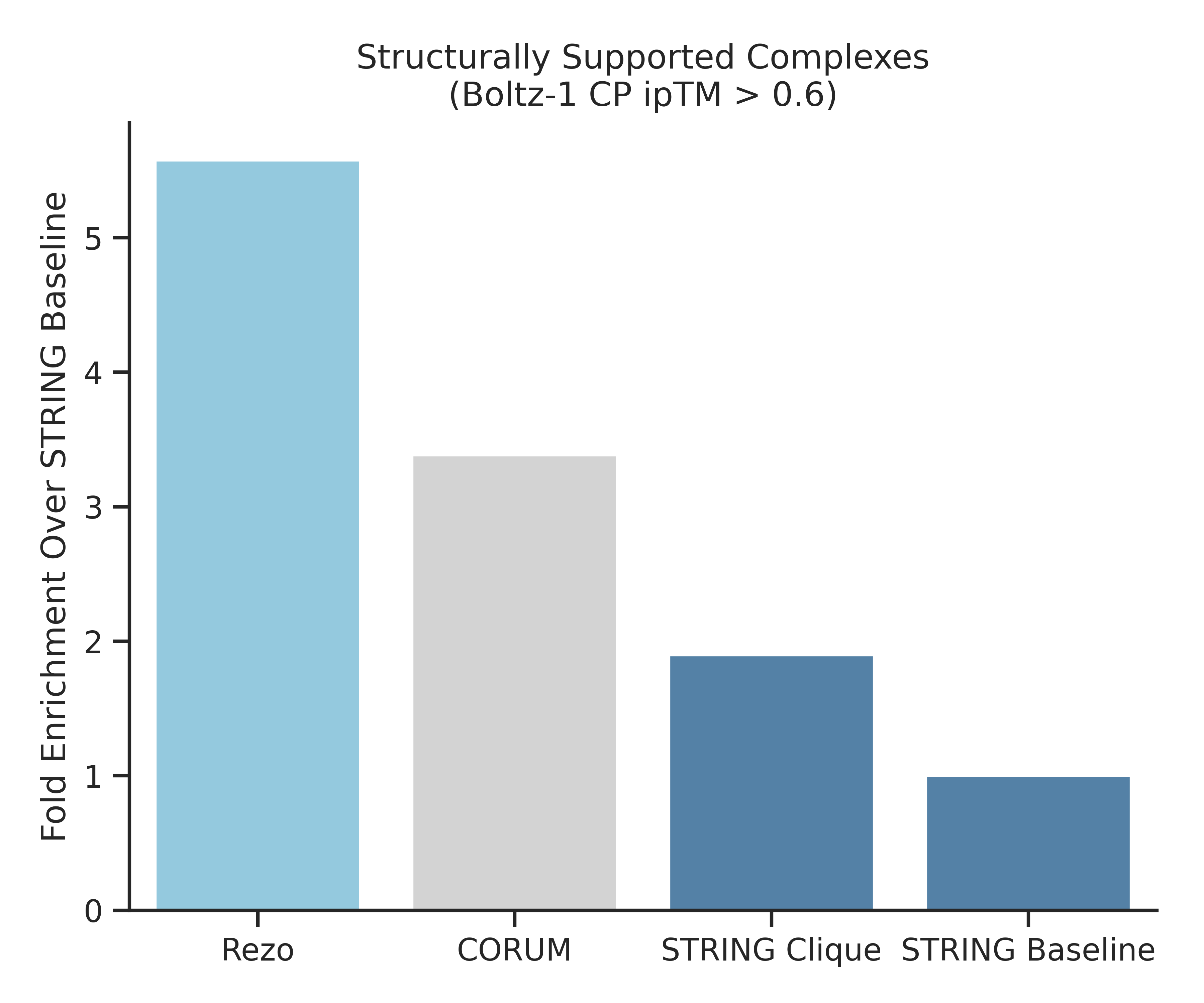}
    \caption{\textbf{Enrichment of different methods for identifying protein complexes, using Boltz-1 CP for structural validation.} From left to right: \textbf{Rezo}, Complexes predicted by Rezo’s Network AI; \textbf{CORUM}, CORUM protein complexes; \textbf{STRING Clique}, High-confidence STRING interactions that form cliques; and \textbf{STRING Baseline}, High-confidence STRING interactions that form connected components but not cliques. To ensure comparability, all evaluated sets were restricted to complexes containing three or four proteins.}
    \label{fig:Rezo_Fig1.png}
\end{figure}

This advance in scale enabled Rezo to structurally evaluate novel protein complexes predicted by their multimodal Network AI model. Predictions from this model, built on proprietary proteomics data, were compared to alternative strategies leveraging public datasets. To establish baselines, STRING interactions were first pruned using a score cutoff of 0.9 prior to complex identification. The resulting complexes were split into two categories: a stricter STRING Clique group (where every protein pair in the complex shared a high-confidence edge) and a STRING Baseline (where the connected proteins did not form a strict clique). In Figure \ref{fig:Rezo_Fig1.png}, the y-axis represents the fold-enrichment of structurally viable complexes compared to the STRING Baseline. Structural viability was defined as achieving a confident interface (ipTM, interface predicted template modeling, score > 0.6) when folded with Boltz-1 CP. Rezo's complexes achieved an enrichment rate greater than 5$\times$ the STRING Baseline, 3$\times$ the stricter STRING Clique set, and even demonstrated stronger support than the manually curated complexes present in CORUM.

\begin{figure}[htbp]
    \centering
    \includegraphics[width=\textwidth]{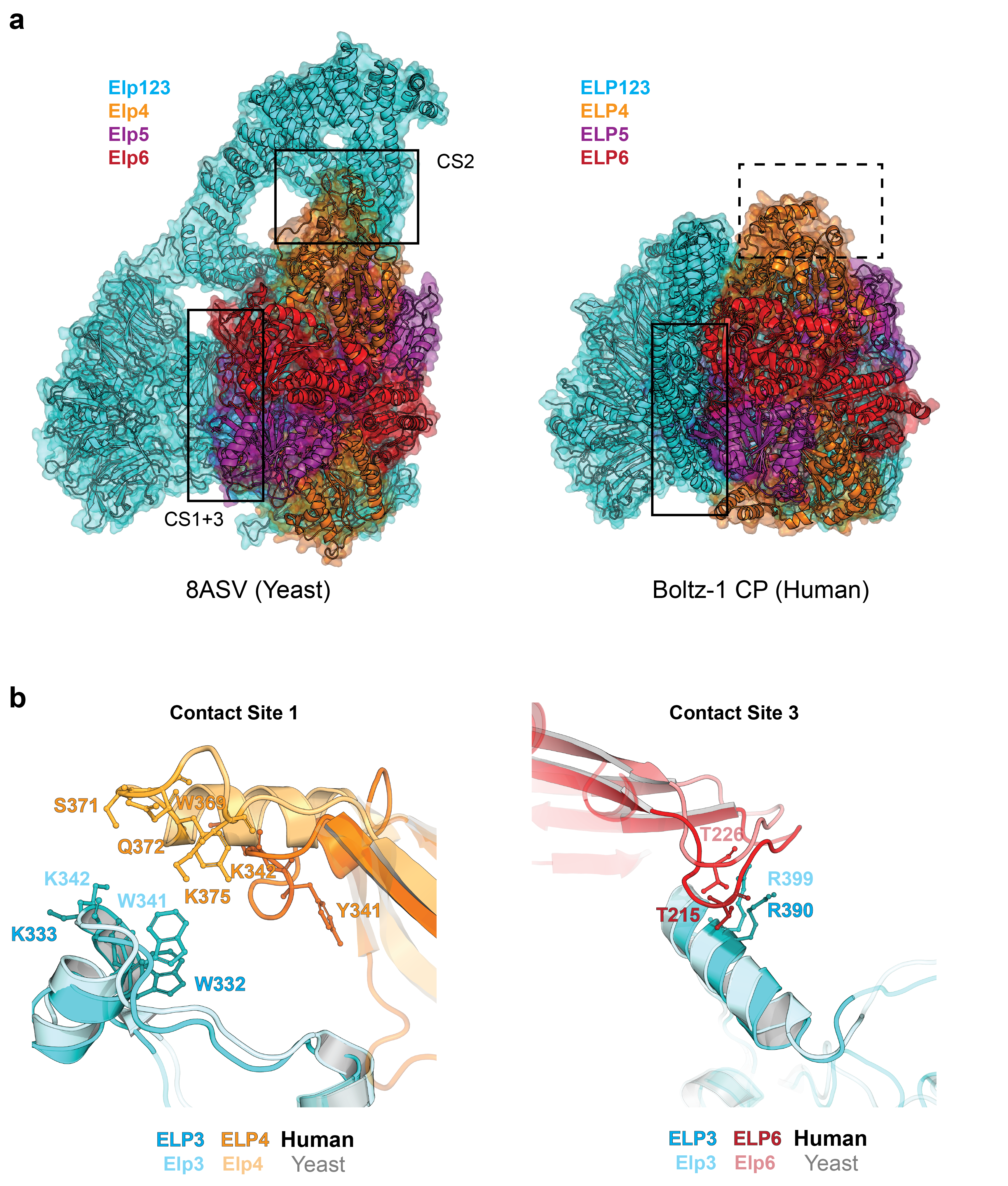}
    \caption{\textbf{Structural comparison of the ELP123-ELP456 complex. (a)} A side-by-side comparison of the complete complex. On the left is the yeast structure (PDB 8ASV~\cite{elp123456}), and on the right is the Boltz-1 CP predicted structure of the human Elongator complex. Outlined sections represent XL-MS contact sites~\cite{xlms} that were either captured (contact sites 1 and 3, CS1+3) or missed (contact site 2, CS2) by the model. \textbf{(b)} A magnified view of the Boltz-1 CP predicted structure, overlaid on 8ASV, appears to recapitulate previously identified interface contact sites (1 and 3)~\cite{elp123456} that were successfully mapped from yeast to human. The remaining contact site 2 resides in a region of low sequence and structural homology between the human (dashed rectangle) and yeast proteins, and likely represents an error in the predicted structure, as detailed in the text.}
    \label{fig:Rezo_Fig2}
\end{figure}

Next, Rezo explored whether any high-scoring complexes had experimental support within the PDB. To test this, they focused on higher-order complexes that were not present in the training data for the Boltz-1 CP model. One representative example is the fully assembled human Elongator complex, which Rezo’s Network AI ranked in the top 5\% of predicted complexes. While individual subunits had previously been resolved, the complete complex was not in the training data. Rezo used Boltz-1 CP to fold the full 4,685-residue, 9-subunit complex. In Figure \ref{fig:Rezo_Fig2}a, the predicted structure is compared to a recent crystal structure of the yeast complex (PDB 8ASV~\cite{elp123456}), which shares a high degree of homology. Encouragingly, the predicted structure correctly identifies two of the previously defined XL-MS contact sites~\cite{xlms} between ELP123 and ELP456 (Figure \ref{fig:Rezo_Fig2}b). 

While the model was able to capture many key structural features in this large complex, the discrepancies in the predicted model provide opportunities for data-driven model improvement. For example, a key difference in the structures is that ELP123 seems to collapse on top of ELP456, missing the remaining contact site (CS2). In particular, ELP456 exists as a homodimer in the resolved structure. This may confound the model’s ability to predict the correct interacting subunit. This example highlights current limitations in the existing model and the potential value of applying Fold-CP at training time. Specifically for large protein complexes, a larger crop size could capture critical structural context, allowing the model to direct ELP1 to the correct ELP4 subunit. Unlocking these higher-order sequences means that providing high-quality training data for this historically undersampled regime will become increasingly critical.

\subsection{Earendil Labs: Accelerating Extreme-Scale Proprietary Foundation Models}

Earendil Labs takes great interest in antibody therapeutics, particularly bispecific and multispecific antibodies. These two types of antibodies can be larger than normal monospecific antibodies with extra binding domains, extra chains, linkers, Fc fusions, or tandem scFvs. In addition, heterogeneous inference batches, where target sizes range from small monomers to massive multi-chain assemblies, have historically crippled the throughput of structural AI pipelines. When massive complexes exceed single-GPU memory capacities, systems must resort to chunking, which creates a significant waste of computing resources in unbalanced batches as many GPUs remain idle, waiting for inference for longer, chunked proteins to complete. 
Even with chunking, the pair representation that has to be materialized in full still limits the size of the complexes that can be studied. By integrating Fold-CP directly into their proprietary, larger-than-normal biomolecular foundation model, Earendil Labs successfully distributed the pair activation tensor across a pool of GPUs. This afforded flexible multi-GPU distribution for single massive targets, maximizing overall throughput and effectively accommodating unbalanced inference batches.

\begin{figure}[H]
    \centering
    \includegraphics[width=0.8\textwidth]{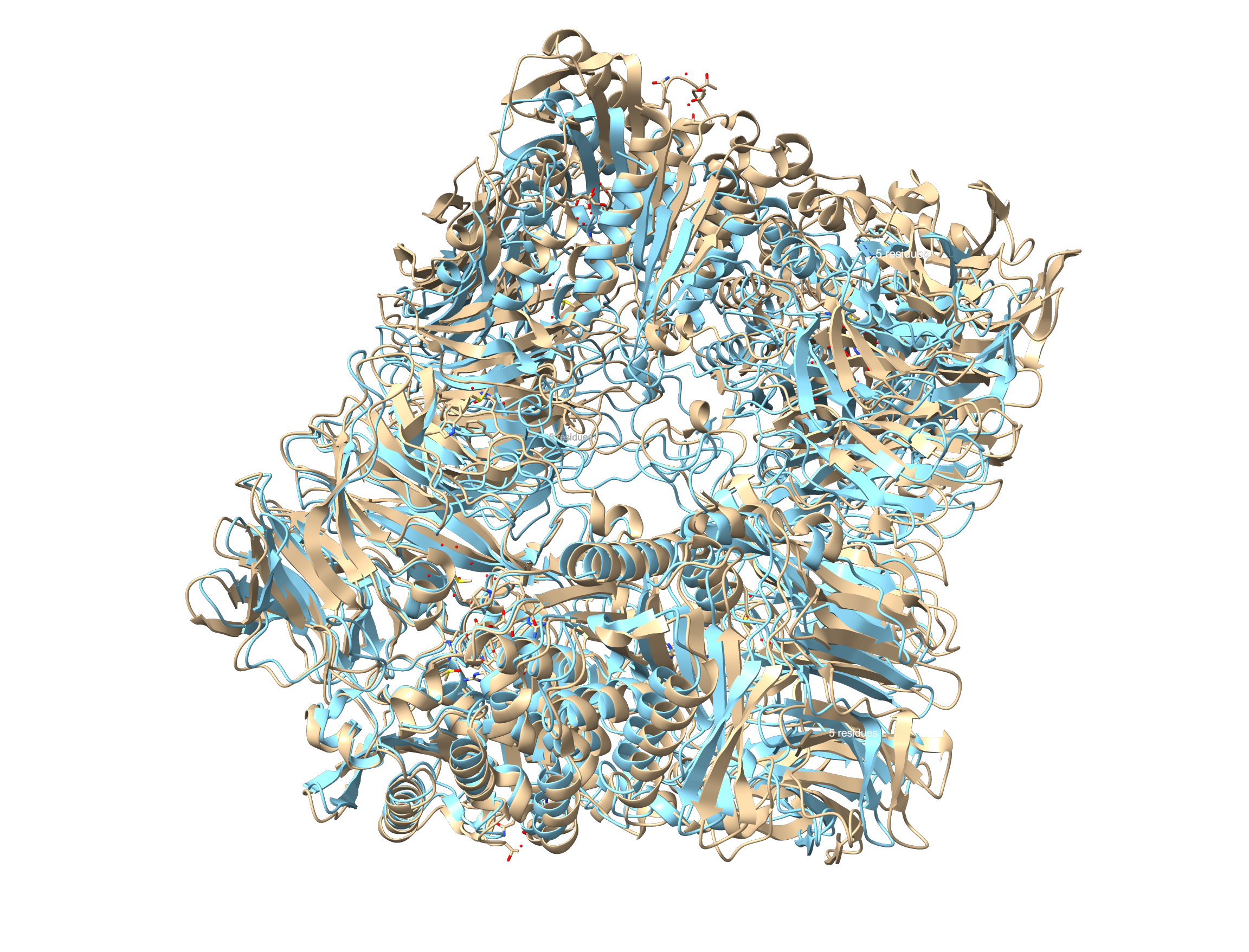}
    \caption{\textbf{Structure prediction of 5YZO homo 4-mer S9 peptidase mutant (S514A) from \textit{Deinococcus radiodurans} R1.}
    Yellow is the ground truth while blue is the predicted version from Earendil's model.
    The prediction was generated in a single inference pass using Fold-CP across 64 NVIDIA H100 80GB GPUs.
    }
    \label{fig:partner_rendering_orgname}
\end{figure}

This integration allowed Earendil to fit sequences of up to 24,000 tokens onto 144 NVIDIA H100 80GB GPUs, simultaneously eliminating the sequential chunking bottleneck and reducing the inference walltime of large complexes by up to two orders of magnitude. Leveraging this extreme-scale inference capability, Earendil successfully folded the 5YZO structure, a homo 4-mer S9 peptidase mutant (S514A) comprising 2,624 amino acids that was previously unable to fit onto a single NVIDIA H100 80GB GPU \autoref{fig:partner_rendering_orgname} with their model. Evaluated across 64 GPUs, the model achieved an RMSD of 4.0. 

By making inference at this scale computationally tractable, Fold-CP provides researchers with the rapid structural insights necessary to interrogate massive therapeutic targets, such as the prolyl oligopeptidase family, across metabolic, neurologic, inflammatory, and other disease areas.

Beyond accelerating inference, the need for a larger context window is equally critical during training. While the short-context-trained model was successful in the case of 5YZO structure as mentioned above, Earendil noted that models often fail to produce coherent structure predictions for larger complexes, even when those targets were present in the training dataset.  

Operating under the assumption that this inability stems from a lack of long-context training, Earendil 
performed training on large crop sizes beyond 768 tokens within their DP-only setup. Validating this hypothesis, they observed an upward trajectory in complex performance, noting that a longer context window results in non-trivial performance improvements for protein-protein complex interactions. Encouraged by this early signal, Earendil anticipates that fully scaled CP training, as enabled by Fold-CP, will empower models to natively produce coherent and correct structures for extreme-scale targets in future work.

\subsection{Proxima: Structurally Resolving the Proximity-Based Interactome}

Proxima's NeoLink cross-linking mass spectrometry platform captures protein interactions as they occur inside living cells to discover and develop proximity-modulating therapeutics. In real world biology, these interactions are rarely simple two-protein encounters but more often large multi-subunit assemblies: signaling hubs, chromatin remodelers, degradation machinery, and regulatory scaffolds that coordinate dozens of proteins across hundreds of thousands of atoms~\cite{marsh2015}. The inter-chain interfaces within these assemblies are precisely the sites relevant to proximity-based drug design: molecular glues, TACs (e.g., RIPTACs, PROTACs), PPI blockers, and related modalities that work by controlling how proteins come together~\cite{Schreiber2021molecularglues, Bekes2022TACs, riptacs}. Structure prediction methods applied to such assemblies have historically been limited due to memory requirements, until the Fold-CP framework introduced in this work. 

Integrated into Proxima's all-atom foundation model Neo, Fold-CP has extended the practical inference limit: structure of thousands of known and novel molecular assemblies at up to 4,000 tokens have been predicted using 4 NVIDIA H100 80GB GPUs. Proxima has built an extensive structurally-resolved PPI and ternary complex dataset, covering over 70\% of the human proteome and diverse species, through its NeoLink + Neo platform \cite{proxima-url}.
Fold-CP extends this to Proxima's experimentally mapped interaction landscape that exceeds current single-GPU sequence limits, adding coverage of the larger multi-subunit assemblies relevant to proximity-based drug development.

\subsection{Boltz-1: Long-Context Inference by a Short-Context-Trained Model}\label{sec:failure_modes}
Many current models are trained in significantly shorter context than their application targets, for example, Boltz-1 training consists of a pretraining stage with crop size of 384 tokens and finetuning with 512 tokens. Model developers have been pushing finetuning context length over time, e.g. Boltz-2~\cite{boltz2} adopts two more stages of finetuning with 640 and 768 tokens respectively, but the efforts to extend the context length significantly beyond has hit the GPU memory wall.  A known failure mode for such short-context-trained models in long-context inference is models predicting unphysical, overlapping chains. This behavior was initially documented in the AF3 \cite{alphafold3} and Boltz-1 publications \cite{boltz1}, with Boltz-1 authors hypothesizing that training with small crop sizes may be the culprit. Our experiments with Boltz-1 CP confirms that this inference behavior is maintained in longer context lengths, in the absence of long context training.

\begin{figure}[H]
    \centering
    \begin{subfigure}{0.35\textwidth}
        \centering
        \includegraphics[width=\textwidth]{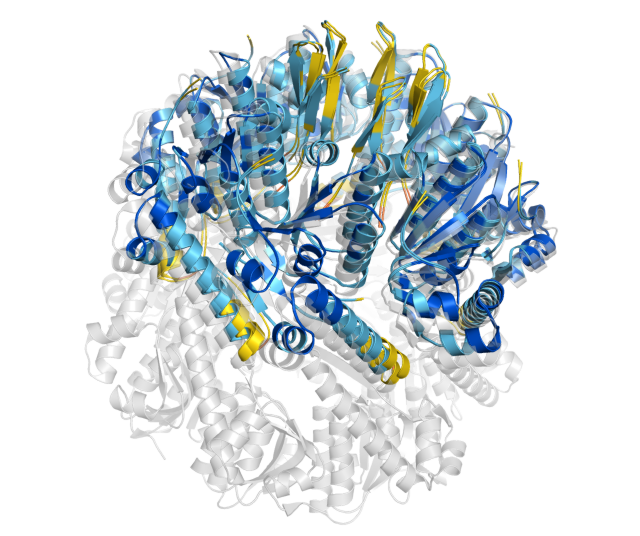}
        \caption{PDB: 2ZL4 }
        \label{fig:2zl4_failure}
    \end{subfigure}
    \begin{subfigure}{0.35\textwidth}
        \centering
        \includegraphics[angle=90, width=\textwidth]{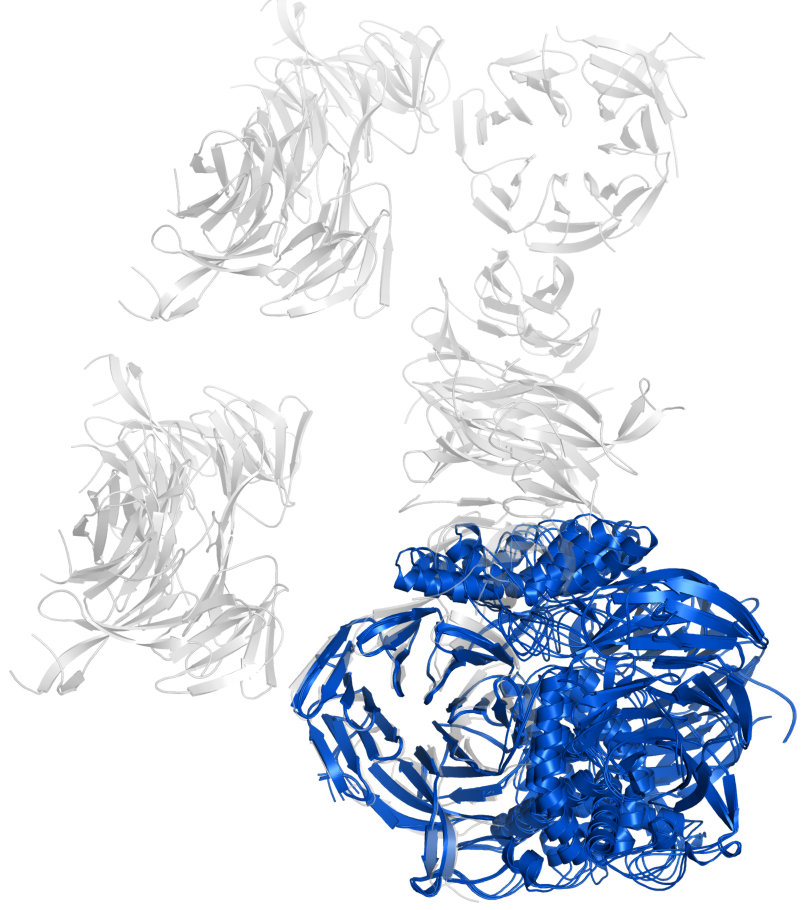}
        \caption{PDB: 8AFY}
        \label{fig:8afy_failure}
    \end{subfigure}
    
    \caption{\textbf{Structural artifacts in long-context homomeric assemblies.} Boltz-1 CP predicted structures exhibit chain overlap and clumping. Experiment structures are in gray. \textbf{Confidence scale (pLDDT):} 90--100 (dark blue), 70--90 (light blue), 50--70 (yellow), 0--50 (orange).}
    \label{fig:common_failure_mode}
\end{figure}

Here we show two exemplary structures that highlight the model failure: a 2,800-token homo-tetradecamer ($D_7$ symmetry, PDB: 2ZL4) and a 4,000-token homo-octamer (PDB: 8AFY), as shown in Figure \ref{fig:common_failure_mode}. For 2ZL4, the expected $D_7$ symmetry collapses into an overlapping $C_7$ configuration. The structure module remains capable of capturing local coordinates within one hemisphere of the experimental structure, while resulting in steric overlap of chains. The confidence model correctly assigns lower scores to loops and less-structured regions, yet it fails to penalize the chain overlaps, maintaining high confidence in such physically impossible regions. Similarly, for 8AFY, the model predicts accurate monomeric folds but erroneously assembles them into a dense "clump" with an average $\text{pLDDT} > 90\%$. This undesirable behavior of confidence metrics invite a renewed attention to designing uncertainty prediction tools that inform about the global 3D organization of biomolecules as well as local structure. 

Although extending steering potentials as introduced in Boltz-1x \cite{boltz1} within the Fold-CP framework remains a viable mitigation strategy, we anticipate that long-context training may also help resolve these artifacts. Allowing the models to learn from large and oblong structures and complex inter-chain interfaces without cropping, may enable models to generalize better in a wider coordinate and stereochemistry space. Finetuning models with increasingly long context may help answer whether this erroneous behavior is solely a short-context-training limitation, or masks other bias introduced by network architecture. The open-source Fold-CP training framework introduced in this work enables the developer community to perform necessary experiments for this exploration.

\section{Conclusion}

The Fold-CP framework represents a convergence of systems engineering and structural biology. By sharding the pair representations efficiently across a 2D GPU device grid and adapting complex algorithms to operate performantly on distributed data, Fold-CP enables unprecedented context lengths in inference and training of AlphaFold-like models, without chunking, or algorithmic approximations to attention mechanisms. Our proof-of-concept implementation on Boltz-1 and Boltz-2 demonstrates that maximum context length scales with the square root of the GPU count. We showcase this capability by performing inference on systems as large as >32,000 tokens on 64 NVIDIA B300 GPUs, bringing massive biomolecular assemblies within the reach of structure prediction modeling.

By enabling holistic predictions of multi-chain assemblies, Fold-CP delivers immediate scientific utility. We demonstrate that the uncropped inference of the 3,605-residue PI4KA lipid kinase complex successfully predicts a novel interaction previously obscured by cropping. Furthermore, our case studies show that researchers can now structurally validate over 90\% of the CORUM database, accelerate proximity-based drug discovery for advanced modalities (e.g., molecular glues and PROTACs), and maximize throughput for extreme-scale foundation models via heterogeneous inference batching.
Through the 4,685-residue human Elongator complex, we show that while many key structural features are identified thanks to increased context, Fold-CP integration also reveals the limitations of the model. 

As Fold-CP opens up predictions at a new context range, a new set of scientific and engineering challenges present themselves that may have been less significant at the small token scale. For example, the square topology requirement of Fold-CP gets increasingly hard to fulfill at very large GPU counts and performance may heavily depend on high speed interconnect like NVIDIA NVLink. Current design opts for a sweet spot for GPU count and code complexity; extension to generic topologies is left for future work.

Another challenge relates to model artifacts at long context range: Large complexes are often assemblies of multiple subunits, for which  the known issue of steric clash of overlapping chains can be significantly limiting. Models may need to be trained or fine tuned on uncropped, large scale complexes to overcome this limitation. These training or fine tuning efforts would highlight another challenge: the data scarcity of ground truth structures for massive token scale. Common data sources as Protein Data Bank are often inherently biased towards smaller, stable, or experimentally truncated structures. For example, approximately 85\% of PDB has fewer than 1,900 tokens \cite{pdb_stats}. Synthetic data generation efforts similar to AlphaFold Database \cite{afdb} and their extensions to longer context may be needed to enable next-generation, CP-native models. By leveraging Fold-CP and high throughput inference tools such as TensorRT\cite{trt}, a new generation of large scale, long context synthetic data may emerge, ready to train next-generation foundation models.

Ultimately, Fold-CP provides the computational engine to process extreme scales, and the case studies with Rezo Therapeutics, Proxima, and Earendil Labs demonstrate the new possibilities that open up with this engine, but scaling the intelligence of biological models will require more effort in several directions including but not limited to large-scale synthetic data generation, novel architectures, and continuous systems engineering tackling multi device communication and single device performance. Combined with these efforts, Fold-CP framework charts a concrete, computationally tractable path toward the holistic simulation of biological systems.

\section*{Code Availability}
A proof-of-concept implementation of Fold-CP on Boltz models is available at \url{https://github.com/NVIDIA-Digital-Bio/boltz-cp}. 

\section*{Acknowledgments}

We thank our partner organizations for their invaluable scientific validation and feedback: Rezo Therapeutics, Proxima, Earendil Labs. We acknowledge the Boltz open-source community \cite{boltz2} and team, including Jeremy Wohlwend and Gabriele Corso for the valuable discussions about Boltz models. We are grateful for fruitful discussions and infrastructure support by several NVIDIA teams: BioNeMo Eng \& BU, HCLS SA, cuEquivariance, cuDNN, Developer Technology, nvFuser, VRDC, Deep Learning Algorithms, Deep Learning Frameworks, Deep Learning Training Performance. 

\bibliography{references}

\end{document}